\documentclass[final,5p,times,twocolumn]{elsarticle}
\usepackage{graphics}
\usepackage{epsfig}
\usepackage{amssymb}
\journal{Radiation Measurements}
\begin{document}

\begin{frontmatter}
\title{ Bulk etch rate measurements and calibrations of CR39 and Makrofol
 nuclear track detectors}

\author[label1]{V. Togo} 
\author[label3]{I. Traor\'e}
\author{for the CERN-EMU-018 Collaboration\corref{cor1}}
\address[label1]{INFN, Sezione di Bologna, Viale C. Berti Pichat 6/2, 
I-40127 Bologna, Italy}
\address[label3]{Physics Department, FAST, University of Bamako, 
BP E2528, Bamako, Mali}
\cortext[cor1]{The {\bf CERN-EMU-018 Collaboration:}  A.B\^{a}, 
S.Balestra, M.Cozzi, G.Giacomelli, R.Giacomelli, 
M.Giorgini, A.Kumar, G.Mandrioli, S.Manzoor, A.R.Margiotta, E. 
Medinaceli, L.Patrizii, V.Popa, I.E.Qureshi, M.A.Rana, Z.Sahnoun, 
G.Sirri, M.Spurio, V.Togo, I.Traor\'e, C.Valieri}

\begin{abstract}
We developed a new method for determining the bulk etch rate velocity 
based on both cone height and base diameter measurements of the etched 
tracks. This method is applied here for the calibration of CR39 and 
Makrofol nuclear track detectors exposed to 158 A GeV In$^{49+}$ and 
Pb$^{82+}$ ions, respectively. For CR39 the peaks corresponding to indium 
ions and their different fragments are 
well separated from $Z/\beta = 7$ to 49: the detection threshold is at REL 
$\sim 50$ 
MeV cm$^{2}$ g$^{-1}$, corresponding to a nuclear fragment with $Z/\beta = 
7$. The 
calibration of Makrofol with Pb$^{82+}$ ions has shown all peaks due to 
lead 
ions and their fragments from $Z/\beta \sim 51$ to 83 (charge pickup). The 
detection threshold of Makrofol is at REL $\sim 2700$ MeV cm$^{2}$ 
g$^{-1}$, corresponding to a nuclear fragment with $Z/\beta = 51$.
\end{abstract}

\begin{keyword}
Calibration; Bulk etch rate; Nuclear track detector.
\end{keyword}

\end{frontmatter}

\section{\bf{Introduction}}
\label{intro}
Nuclear Track Detectors (NTDs) are commonly used in many branches of 
science and technology\cite{Fleischer,Durrani}. The 
isotropic 
poly-allyl-diglycol carbonate polymer, commercially known as CR39® is the most sensitive NTD; also Makrofol® and Lexan® polycarbonates are largely 
employed. More than 4000 m$^2$ of CR39 detectors were used in the MACRO and SLIM experiments devoted to the search for new massive particles in the cosmic 
radiation (magnetic monopoles, nuclearites, Q-balls) 
\cite{Ambrosio2002a,Ambrosio2002b,Balestra2008,Cecchini2005a,Cecchini2005b,Derkaoui,Giacomelli2007,Manzoor2007,Medinaceli,Sahnoun}. 
Several experiments are going on in different fields which require an accurate detector calibration \cite{Uchihori,Kodaira}.
A nuclear track detector can record the passage of ionizing particles or 
any highly charged particle via their Restricted Energy  Loss  (REL).  The  
latent  damage trail formed in NTDs may be enlarged by a suitable chemical 
etching and made visible under an optical microscope. The latent track 
develops into a conical-shaped etch-pit (Fig.1) when the etching velocity 
along the particle trajectory ($v_{T}$) is larger than the one for the 
bulk etching of the material ($v_{B}$)\cite{Nikezic}.
The sensitivity of NTDs crossed by particles with constant energy loss can 
be characterized by the ratio $p = v_{T}/v_{B}$ (reduced etch rate) which 
may be determined by measuring the bulk etch rate $v_{B}$ and either the 
etch-pit diameter or the etch-pit height. Two methods have been used to 
determine $v_{B}$. The first is the common one based on the mean thickness 
difference determination before and after etching. The second method that 
is the main aim of this paper, is based on both cone height and base diameter 
measurements of the etched tracks.

\section{\bf {Experimental}}
\subsection{\it {Detectors and exposure}}
A stack composed of Makrofol and CR39 foils of size 11.5 x 11.5 cm$^{2}$ 
with a 1 cm thick lead target was exposed to 158 A GeV Pb$^{82+}$ ions in 
1996 (Pb96); a second stack with a 1 cm thick aluminium target was exposed 
to 158 A GeV In$^{49+}$ ions in 2003 (In03); both exposures were 
performed 
at the CERN-SPS, at normal incidence and a total ion density of 
$\sim$ 2000/cm$^{2}$. 

\begin{figure}[ht]
{\resizebox*{!}{5cm}{\includegraphics{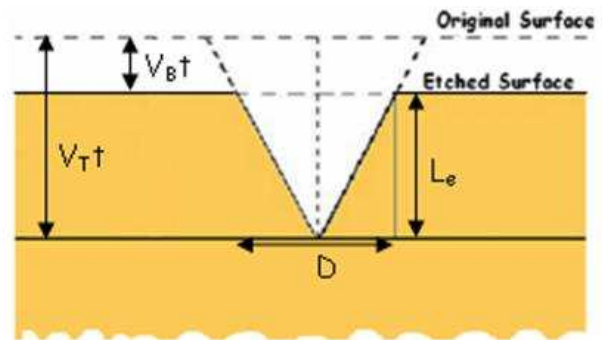}}}
\begin{quote}
\caption{\small Sketch of an "etched track" for a normally incidt ion 
in a nuclear track detector.} 
\label{fig:1}
\end{quote}
\end{figure}

\begin{figure}[h]
{\resizebox*{!}{6.5cm}{\includegraphics{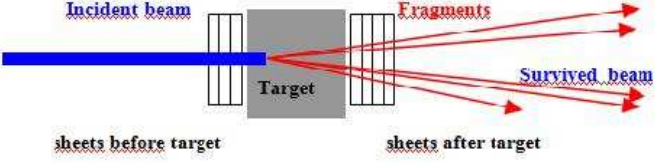}}\par}
\vspace{-1cm}
\begin{quote}
\caption{\small Exposure set-up for the calibration of CR39 and Makrofol 
NTDs.} 
\label{fig:2}
\end{quote}
\end{figure}

The detector foils downstream of the target recorded the beam ions as well as their nuclear fragments \cite{Cecchini2008,Giorgini,Togo}, see Fig2.
The CR39 polymer sheets used in the present study were manufactured by 
Intercast Europe Co., Parma, Italy where a specific production line was 
set up in order to achieve a lower detection threshold, a higher 
sensitivity in a larger range of energy losses, a high quality of the 
post-etched surfaces after prolonged etching \cite{Ahlen91,Fantuzzi,Patrizii,Vilela}. The Makrofol 
detectors were manufactured by Bayer A.G., Germany. They have a thickness 
of 500 $\mu$m. The CR39 thickness is either 700 $\mu$m or 1400 $\mu$m. All 
detector sheets were covered by a 30 $\mu$m plastic film to protect them 
from 
exposure to ambient radon; the protective layers were removed before etching. 
\subsection{\textit{Etching conditions}}
After exposures, two CR39 foils and two Makrofol foils located after the target were etched. The etchant used were 6N NaOH or 6N KOH at temperatures in 
the range from $45 ^\circ$C to $70 ^\circ$C and different etching times. 
The addition of ethyl alcohol in the etchant \cite{Matiullah} improves the 
etched surface quality, reduces the number of surface defects and background tracks, increases the bulk etching velocity, 
speed up the reaction, but raises the detection 
threshold \cite{Manzoor2007,Balestra2007}. It results that 
6N NaOH + 1\% ethyl alcohol at $70^\circ$C for 40 h and 6N KOH + 20\% 
ethyl 
alcohol at $50^\circ$C for 8 h represent the optimum etching condition 
for CR39 and Makrofol, respectively.
The etching was performed in a stainless steel tank equipped with internal 
thermo-resistances and a motorized stirring head. 
In order to keep homogeneous the solution and to avoid that etched products 
deposit on the detector surfaces, a continuous stirring was applied during 
etching. The temperature was stable to within $\pm 0.1$ $^\circ$C.
For CR39 detectors, etch-pit base diameters and heights of In ions and 
their fragments were measured with a Leica optical microscope. 
In Makrofol, Pb ions and their high Z fragments ($Z>77$) made 
through-holes 
in the detector sheets; thus the cone length $L_{e}$ was measured only for 
fragment tracks that have sharp etch-cone tips (no holes). Nuclear fragments 
with Z charges $78 \leq Z \leq 82$ were identified by etching another 
Makrofol 
sheet from the same stack in the same conditions but for only 5 h.
\subsection{\textit{Bulk etch rate from the thickness-changing measurements.}}
For the determination of $v_{B}$, the thickness of the detectors was 
measured in 25 selected points with an electronic micrometer of 1 $\mu$m 
accuracy. The average bulk etch velocity is $v_{B} = \Delta x/ 2 \Delta t$ 
, where $\Delta x$ is the mean 
thickness difference after a $\Delta t$ etching time. For CR39, at etching 
time 
shorter than 10 h the thickness is affected by detector swelling 
\cite{Ahlen93,Kumar,Malik}. For Makrofol 
no significant swelling effect was observed.
\subsection{\it {Bulk etch rate from the cone height and base 
diameter measurements}}
For relativistic charged particles the track etch rate $v_{T}$ can be 
considered constant. For normally incident particles, the measurable quantities are the cone base diameter D, and the height $ L_{e}$ 
\cite{Balestra2007} see Fig1. The following relations hold:

\begin{equation}
L_{e} = (V_{T}-V_{B}) t
\end{equation}

\begin{equation}
D = 2V_{B}t\sqrt{\frac{(V_{T}-V_{B})}{(V_{T}+V_{B})}}
\label{eq:2}
\end{equation}

From the above relations, the following quadratic equation in $v_{B}$ is 
obtained:
\begin{equation}
\left(\frac{L_{e}}{t}\right) V_B^{2}-\left(\frac{D^{2}}{2t^{2}}\right) V_{B}-\left(\frac{D^{2}L_{e}}{4t^{3}}\right)=0
\label{eq:3}
\end{equation}

The real solution for $v_{B}$ is:
\begin{equation}
V_{B}= \frac{D^{2}}{4tL_{e}} \left[1+\sqrt{ \left(1+\frac{4L^{2}_{e}}{D^{2}}\right)} \right]
\label{eq:4}
\end{equation}

From Eq.(1) the track etch rate $v_{T}$ can be written as:
\begin{equation}
V_{T}= V_{B}+\frac{L_{e}}{t}
\label{eq:5}
\end{equation}

From Eqs. (1) and (2), the reduced etch rate follows:
\begin{equation}
P = \frac{V_{T}}{V_{B}} = 1 + \frac{L_{e}}{V_{B}t} = \frac{1+(\frac{D}{2V_{B}t})^{2}}{1-(\frac{D}{2V_{B}t_{}})^{2}}
\label{eq:6}
\end{equation}

We may thus determine the bulk etch rate $v_{B}$ and the reduced etch 
rate 
p by measuring the track parameters $L_{e}$ (measured with a precision of 
$\sim 1$ $\mu$m) and D (precision of 0.5 $\mu$m).

\begin{table*}
\begin{center}
{\small
\begin{tabular}
{|c|c|c|c|c|}\hline
{\bf Detector (beam) } & {\bf Z range} & {\bf Etching conditions} & 
{\bf $V_{B}$ new method} & {\bf $V_{B}$ thickness method} \\
 & & & {\bf ($\mu$m/h)} & {\bf($\mu$m/h) }\\
 \hline
CR39(In03) & 44-49 & 6N NaOH+1\% alcohol, $70^\circ$C, 40h. & 1.25 $\pm$ 
0.01 & 1.15 $\pm$ 0.03 \\
CR39(Pb96) & 75-80&6N NaOH $70^\circ$C, 30h. & 1.10 $\pm$ 0.02 &1.15 $\pm$ 
0.03 \\
CR39(Pb96) & 78-82&6N NaOH $45^\circ$C, 268h. & 0.16 $\pm$ 0.01 &0.17 $\pm$ 
0.03 \\
Makrofol(Pb96) & 75-78&6N KOH+20\%alcohol, $50^\circ$C, 8h. & 3.44 $\pm$ 
0.05 &3.52 $\pm$ 0.13 \\
\hline
\end{tabular}
}
\end{center}
\caption {Bulk etch rates $v_{B}$ for CR39 and Makrofol NTDs obtained with 
the new and  the thickness-changing methods.
The errors are statistical standard deviations of the mean. The different values of $ v_{B}$ for Pb96 in rows 3 and 4 are due to the different etching 
temperatures.} 
\label{table:1}
\end{table*}

Relations  (4) and (6) were tested with relativistic Pb and In ions and their nuclear fragments. We selected only tracks for which precise measurements of the cone height and diameter could be performed (for example we cannot measure the track cone heights for low  Z  fragments, for which  the microscope  image  may  be  affected  by  shadow effects). 
Then, using Eq. (4) we computed the bulk etch rate for CR39  and  Makrofol.   
Batches of measurements were made by different operators, and the average $v_{B}$''s and their statistical standard deviations were computed, see  Table 
1. By  
this  method  we  obtain  $v_{B}$  values  with  accuracies  of $\pm$ 0.01 to 0.05 $\mu$m/h. The $ v_{B}$ values obtained for the same foils using 
detector thickness measurements are also given.
 \section{\textbf{Calibrations}}
In the following sections calibration data based on the new determination of the bulk etch rate for Makrofol exposed to Pb$^{82+}$ ions and CR39 exposed 
to In$^{49+}$ ions are reported. 
The calibration is given as a relation between the reduced etch rate p and the particle restricted energy loss (REL) \cite{Baiocchi,Cecchini96}. 

 \begin{figure}[h]
{\resizebox*{!}{7.5cm}{\includegraphics{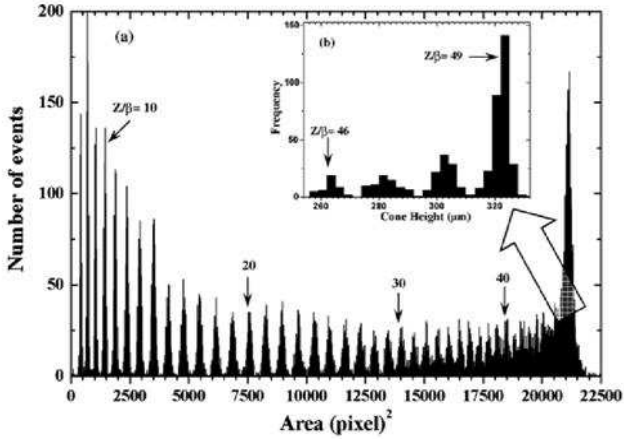}}\par}
\vspace{-1cm}
\caption{\small Base area distribution of etched cones in 
CR39 from 158 A GeV In$^{49+}$ ions and their fragments (averages 
of two front face measurements). In the insert are shown 
the cone height distributions for $46 \leq Z/\beta \leq 49$. Etching 
conditions: 6N NaOH+1\% ethyl alcohol, $70^\circ$C, 40 h.} 
\label{fig:3}
\end{figure}

\begin{figure}[h]
{\resizebox*{!}{7.5cm}{\includegraphics{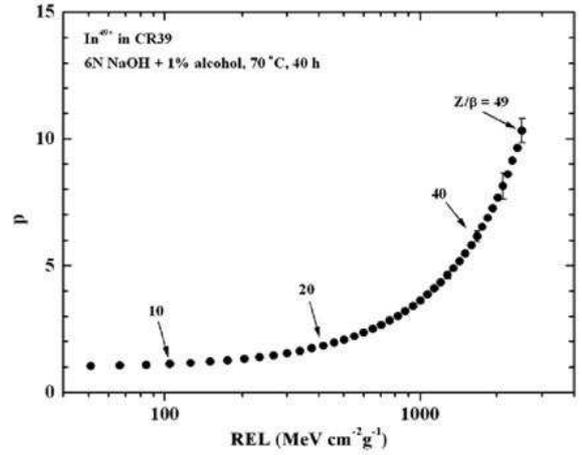}}\par}
\vspace{-1cm}
\begin{quote}
\caption{\small p versus REL for CR39 exposed to relativistic indium 
ions using $v_{B}$ evaluated with the new method. Typical statistical 
standard 
deviations are shown at  $Z/\beta = 40, 45, 49$; for $Z/\beta = 37$ the 
errors 
are smaller than the black points.} 
\label{fig:4}
\end{quote}
\end{figure}

\begin{figure}[h]
{\resizebox*{!}{7.5cm}{\includegraphics{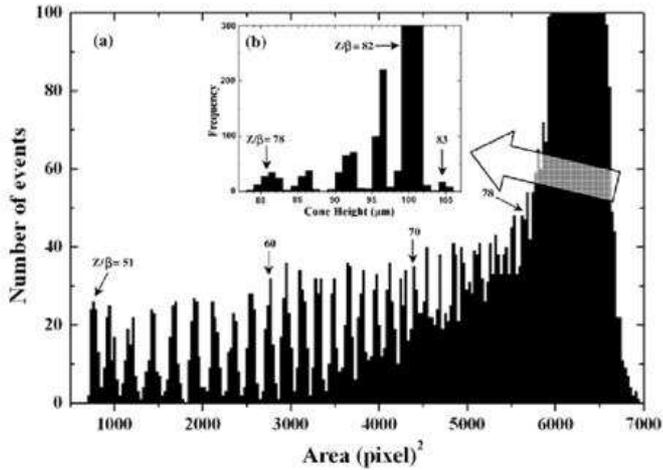}}\par}
\vspace{-1cm}
\begin{quote}
\caption{\small Base area distribution of etched cones in Makrofol from 158 
A GeV Pb$^{82+}$ ions and their fragments (averages of two front face 
measurements). Etching conditions: 6N KOH+20\% ethyl alcohol,$50^\circ$C, 
8 h. 
In the insert are shown the cone height distribution for 
$78 = Z/\beta = 83$. Etching conditions: 6N  KOH+20\% ethyl alcohol, $50^ 
\circ$C, 5 h.} 
\label{fig:5}\end{quote}
\end{figure}

\begin{figure}[h]
{\resizebox*{!}{7.5cm}{\includegraphics{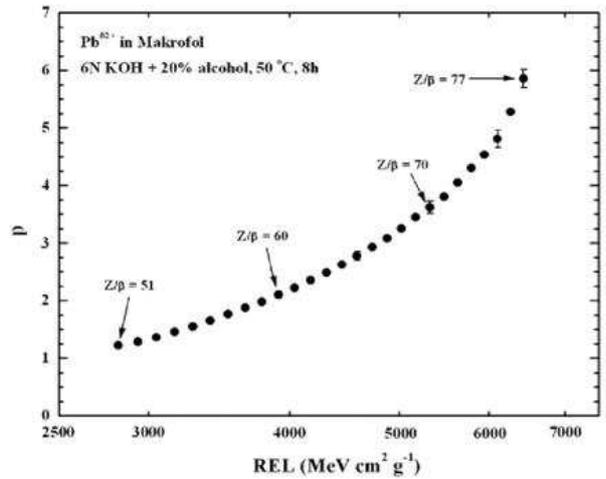}}\par}
\vspace{-1cm}
\begin{quote}
\caption{\small p versus REL for the Makrofol detector exposed 
to relativistic Pb ions using $v_B$ evaluated with the new method. Typical 
statistical standard deviations are shown at  $Z/\beta = 70, 75, 77$; for 
$Z/\beta 
= 67$ the errors are smaller than the black points.} 
\label{fig:6}\end{quote}
\end{figure} 

\subsection{\textit{Calibration of the CR39 detector}}
     Fig.3 shows the etch-pit base area distribution for indium ions and their fragments in CR39 measured with the Elbek automatic image analyzer system 
\cite{Noll} ; averages were computed from measurements made on the ''front 
sides'' of two detector sheets. The peaks are well separated from 
$Z/\beta 
\sim 
7$  to  45; the charge resolution for the average of two measurements is 
$\sigma_{z} \sim $0.13e at $Z/\beta \sim  15$. The charge resolution 
close to 
the 
indium peak ($Z\sim 
49$) can be improved by measuring the heights of the etch-pit cones 
\cite{Giacomelli98}. The height of 1000 etch-cones with diameter 
larger than 48 
$\mu$m (corresponding to nuclear fragments with $Z > 45$) were measured 
with an accuracy of $\pm$ 1 $\mu$m with a Leica microscope coupled to a 
CCD camera 
and a video 
monitor. The corresponding distribution is shown in the insert in  Fig.3; each of the four peaks is well separated from the others, and a charge can be 
assigned to each one \cite{Cecchini93}.
For each detected nuclear fragment from $Z/\beta = 7$ to 48 and indium ions 
(Z = 49) we computed the REL and the reduced etch rate $ p = v_{T}/v_{B}$ 
using  Eq. 
(6); p versus REL is plotted in Fig.3; the CR39 detection threshold is at REL $\sim 50$ MeV cm$^{2}$ g$^{-1}$ (corresponding to a relativistic nuclear 
fragment 
with Z$\sim 7$).

\subsection{\textit{Calibration of the Makrofol detector}}
 Fig.5 shows the base area distribution for the average of two measurements of Pb ions and their fragments in Makrofol; averages were computed from 
measurements on the front sides of two detector foils. The peaks are well separated from $Z/\beta \sim 51$ to $\sim 77$ (the charge resolution is 
$\sigma_{z} \sim 0.18e$  at  
$Z/\beta\sim 55$). The charge resolution close to the Pb peak (Z = 82) was improved by measuring the heights of the etch-pit cones. The heights of 4000 
etch 
cones with base diameter larger than 47 $\mu$m were measured \cite{Manzoor2005}; the corresponding distribution is shown in the insert in Fig.5; 
each 
peak is well separated from the others, and a charge was assigned to every peak.
\section{\bf{Discussion and Conclusions}}
We have studied the response of CR39 and Makrofol NTDs to relativistic indium and lead ions and their fragments using measurements of the etch-pit base area and of the heights of the etch-pit cones. The bulk etch rate measurements using the "new method" yields slightly smaller uncertainties than the ''standard method'' (change in thickness). 
The distributions of the peaks for the primary ions and for their fragments are given in Figs.3 (for CR39) and 5 (for Makrofol). At low $Z/\beta$ the 
measurements of the etch-pit base area are adequate; it is seen that for high $ Z/\beta$, the base area distribution does not give well separated peaks, 
while by cone height measurements the peaks are well separated (see the inserts in Figs. 3 and 5). 
The reduced etch rate p (computed with the new method) plotted versus REL covers a large $Z/\beta$ range for both detectors, Figs.4 (for CR39) and 6 (for 
Makrofol).
\section*{\bf{Acknowledgments}} 
We thank the CERN SPS staff for facilitating the Pb and In beam exposures. We acknowledge many colleagues for their cooperation and technical advice. We gratefully acknowledge the contribution of our technical staff in Bologna. We thank INFN and ICTP for providing fellowships and grants to non-Italian citizens.

\end{document}